 \documentclass[final,5p,times,twocolumn]{elsarticle}
\usepackage{graphicx}
\usepackage{amssymb}
\usepackage{gensymb}
\usepackage{natbib}
\usepackage{mathtools}
\usepackage{booktabs}
\usepackage{lineno}

\begin{document}

\begin{frontmatter}
\title{Geometrical and energy scaling in the pulsed laser deposition plasma during epitaxial growth of FeSe thin films}



\author[label1]{S. B. Harris \corref{cor1}}
\ead{sumner@uab.edu}

\author[label1]{K. L. Kopecky}

\author[label1]{C. W. Cotton}


\author[label1]{and R. P. Camata}

\address[label1]{Department of Physics, University of Alabama at Birmingham, Birmingham, Alabama 35294, USA}
\cortext[cor1]{Corresponding author}

\begin{abstract}
Pulsed laser deposition (PLD) is a versatile technique for growing epitaxial heterostructures of a wide variety of novel materials combinations. Achieving low-defect layers with atomically sharp interfaces by PLD requires careful management of crystal growth conditions. Control over the flux of depositing particles, their kinetic energy, and the substrate temperature, is generally sufficient to obtain high-quality single crystal epitaxy in vacuum. In this article, we show that measurements of plasma parameters such as the electron temperature, the electron density, and the Mach number of the plasma fluid expansion may provide additional insights into the tuning of the crystal growth and processing environment of PLD. We report Langmuir probe measurements during growth of FeSe thin films on (100)-oriented SrTiO$_3$. We discuss two distinct plasma regimes that are accessible by KrF laser ablation of FeSe when pulse energies in the 20-630 mJ range are used. The two regimes can be characterized by the relative number of photons to absorbing centers in the laser-plasma interaction volume. Thin films grown under the conditions created by these two distinct plasma regimes are analyzed by x-ray diffraction and x-ray reflectivity and their epitaxial configurations correlated to the PLD conditions associated with these plasmas.

\end{abstract}

\journal{arXiv}
\end{frontmatter}

\section{Introduction}
\label{S:1}
The combination of different materials in heterostructures often enables phenomena that are impossible to achieve in a single compound. A now-classic example is the interface-enhanced superconductivity of monolayer FeSe on (001)-oriented SrTiO$_3$ (STO). The observed superconducting gap in this system is substantially larger than in bulk FeSe due to a combination of electron doping and electron-phonon interactions at the FeSe/STO interface \cite{Song2019}. The gap increase correlates with a rise in the superconducting critical temperature from 8 K in bulk FeSe to possibly as high as 109 K in monolayer FeSe/STO \cite{Bozovic2014}. 
Heterostructures derived from this FeSe/STO system are useful for studying interface-enhanced processes \cite{Zhang2017}, magnetic/superconducting proximity effects \cite{Kim2012}, and tuning of quantum properties via electrostatic gating \cite{Shiogai2016}. Pulsed laser deposition (PLD) is a versatile technique for growing such heterostructures either as stand-alone thin films on a variety of substrates or as designed multilayered and superlattice configurations. It is well known that PLD growth of epitaxial FeSe layers with low defect density and atomically sharp interfaces requires careful control of deposition rate, kinetic energy of depositing particles, and substrate temperature. Correlations between ranges of these PLD parameters and resulting FeSe film characteristics have been reported in recent years for variety of substrates \cite{feng2019highthroughput, Harris2019, Feng2018, Hiramatsu2014}. 

In this paper, we further characterize the environment in which PLD of FeSe takes place, by probing the laser plasma linked with different epitaxial growth conditions of FeSe on (001)-oriented STO. Measurements of the plasma parameters at the location of film growth were carried out as a function of laser fluence and laser spot size, using Langmuir probes. We evaluate how trends in the electron number density, electron temperature, Debye length, and Mach number of the plasma fluid expansion correlate with the governing PLD parameters associated with the laser plume, namely the kinetic energy and flux of depositing species. Insights gained from measurements of the laser-produced plasma that mediates crystal growth in PLD may provide additional guidance and control in layer-by-layer epitaxy of FeSe-based systems, and other emerging quantum heterostructures.  

\section{Experimental Details}
\label{S:2}
\begin{figure}[t!]
    \centering
    \includegraphics[width=\linewidth]{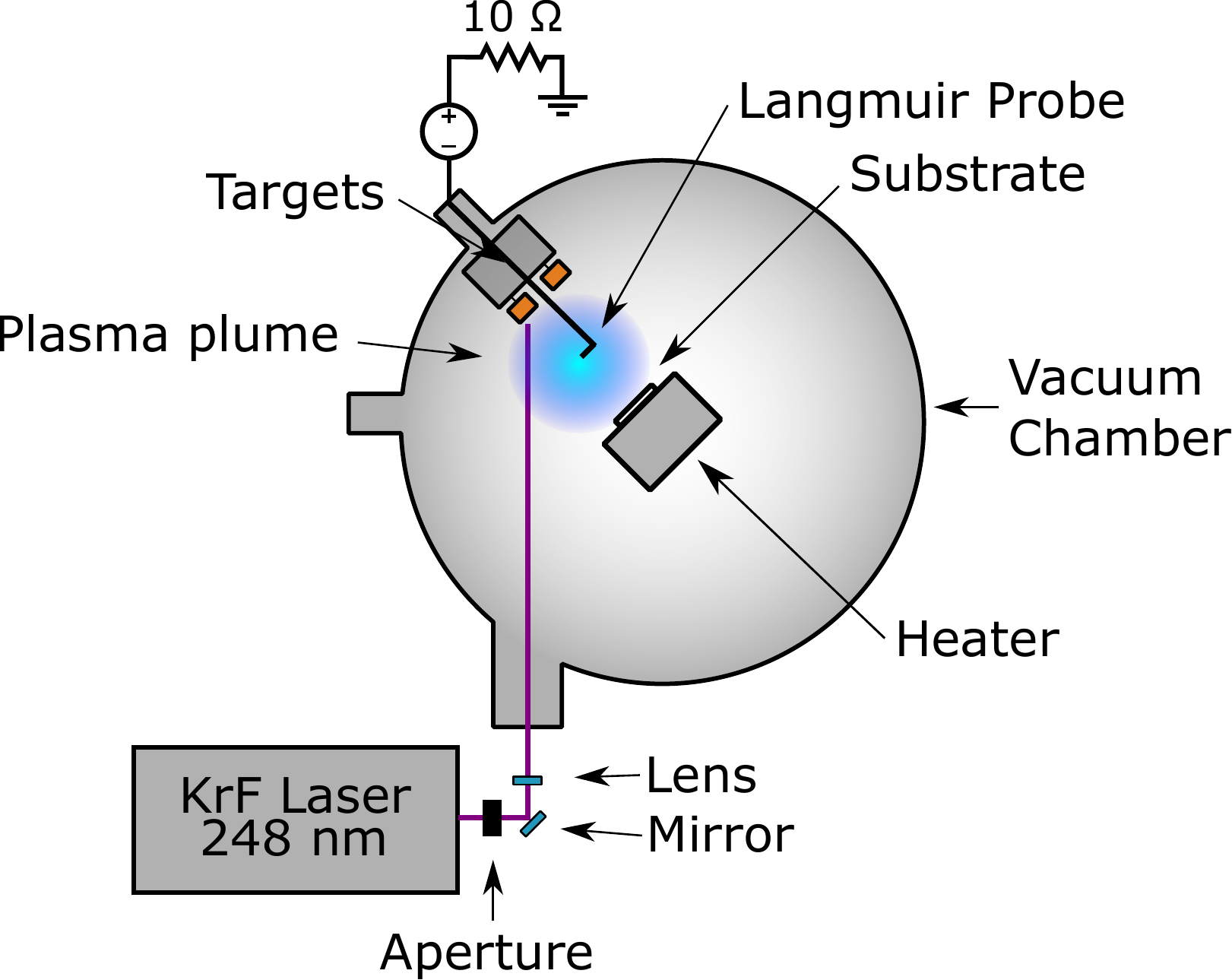}
    \caption{Schematic of the pulsed laser deposition system, including Langmuir probe. A 10 $\times$ 10 mm planar probe is used to acquire ion TOF data, whereas $I$-$V_p$ characteristics are studied with a 101.6-$\mu$m radius (32 AWG) cylindrical probe of length 4.5 mm. Each probe is attached to a micrometric linear translation feedthrough that allows its placement at distances between zero and $5.500 \pm 0.005$ cm from the target. A rectangular aperture of variable size is placed before the focusing lens to change the spot size on the target while maintaining a fixed energy density.}
    \label{experiment_schematic}
\end{figure}

Experiments were carried out in the PLD system shown schematically in Fig. \ref{experiment_schematic}, whose configuration has been described in detail elsewhere \cite{Harris2019}. Briefly, the focused beam of a KrF excimer laser (Lambda Physik LPX 305i), which has pulse duration of 20 ns and wavelength of 248 nm, is used to ablate a rotating 1-inch diameter FeSe target, at an incidence angle of 45$\degree$, in vacuum at a background pressure of $3\times 10^{-6}$ Torr. The laser spot size on the target was controlled using rectangular apertures placed in the beam path between the laser output window and the focusing lens (Fig. \ref{experiment_schematic}). Imaging apertures of different sizes on the target, changes the size of the focused spot, with the laser fluence remaining approximately the same. The laser fluence itself was varied by adjusting the pulse energy of the laser with a fixed aperture size.

The FeSe ablation target was synthesized from metal powders of the elemental precursors with the nominal stoichiometry FeSe$_{0.97}$ using the technique reported by Feng et al. \cite{Feng2018}. This target stoichiometry produces the highest superconducting critical temperature in thick FeSe films. Properties of the plasma produced from the target were measured with Langmuir probes of two different geometries. A cylindrical probe consisting of a 101.6-$\mu$m (32 AWG) radius, 4.5-mm long platinum wire was used to measure plasma parameters. By stepping the probe bias, cylindrical probe theories permit extraction of the electron temperature ($T_e$) and the electron number density ($n_e$) from the probe current vs. bias voltage characteristics ($I$-$V_{p}$) \cite{Weaver1999}. Alternatively, a 10 $\times$ 10 mm square planar probe, composed of a thin molybdenum foil backed with alumina, was used under negative bias for measurements of the time of flight (TOF) distribution of ions in the plasma. Each probe was attached to a linear translation vacuum feedthrough in order to be positioned at distances between zero and $5.500 \pm 0.005$ cm from the target along the axis of symmetry of the ablation plume. The probes are connected to a bias voltage source and the collected current is determined as a function of time ($I$-$t$) from the voltage drop across a 10 $\Omega$ resistor, connected to ground. An oscilloscope was used to average the signal produced by at least 30 laser pulses for each value of probe bias. The target was cleaned with 1000 laser pulses before data collection began.

In order to probe the laser plasma associated with different epitaxial growth conditions, PLD parameters were adjusted to achieve FeSe thin films deposited under distinct laser plasma regimes. All films were grown at 500$\degree$C on TiO$_2$-terminated (001)-oriented STO substrates prepared using a DI water etching and annealing process \cite{Connell2012}.
Thin films were subjected to x-ray diffraction (XRD), x-ray reflectivity (XRR), and reciprocal space mapping (RSM) measurements, performed on a Panalytical Empyrean x-ray diffractometer with a Cu anode ($K_{\alpha1}$ = 1.540598
\AA, $K_{\alpha2}$ = 1.544426 \AA) using x-ray optics that attenuate K$_\beta$.
Incident divergence and anti-scatter slits were 1/8$\degree$ and 1/16$\degree$ to create a quasi-parallel beam. The PIXcel3D detector was used with a 7.5 mm anti-scatter slit in 1D scanning line mode and 1D frame based mode for XRD and RSM, respectively. XRR used a 1/16$\degree$ anti-scatter slit and receiving slit mode (0.055 mm active area).

\begin{figure*}[!th]
    \centering
    \includegraphics[width = \linewidth]{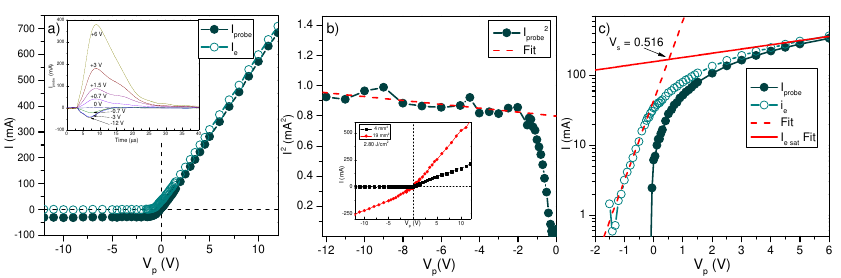}
    \caption{a) $I$-$V_p$ characteristic before (solid circles, $I$) and after (open circles, $I_e$) ion current subtraction. The inset shows the time dependence of probe current for several different values of bias. b) $I^2$-$V_p$ in the ion saturation region is fitted with OML theory to determine ion/electron density. Inset shows $I-V_p$ characteristics collected at 15 mm from the ablation target. For lower values of the spot  area (e.g., 4.0 mm$^2$), OML theory clearly remains valid when the probe is close to the target but the ion current contribution becomes too significant as for large spot sizes such as 19 mm$^2$. c) $I$-$V_p$ characteristic before and after ion current subtraction on a semi-log scale, around the space potential. The linear portion is extended to lower $V_p$ after ion current subtraction and is fitted to determine $T_e$. The electron saturation is also fitted to find $V_s$ at the intersection with of the two lines.}
    \label{typical_IV}
\end{figure*}


\section{Langmuir Probe Data Analysis}
\label{S:3}

Obtaining reliable values for the plasma parameters from Langmuir probe data requires careful considerations for each probe geometry. For cylindrical probes, which allow measurement of the electron temperature, a central point is the need to separate the contributions from the ion current ($I_i$) and the electron current ($I_e$) to the $I$-$V_p$ dependence. A typical cylindrical probe $I$-$V_p$ curve of the plasma produced during FeSe ablation is shown in Fig. \ref{typical_IV}a. When the probe bias ($V_p$) is sufficiently more negative than the space potential $V_s$ the probe collects the ion saturation current $I_{isat}$. Conversely, when $V_p$ is much more positive than $V_s$ the electron saturation current ($I_{esat}$) is collected, which represents the random thermal electron current to the probe. In between these extremes, the probe current involves contributions from both, ions and electrons. A good starting point to sort out the electron and ion contributions is to first deduce the electron current to the probe from thermodynamic considerations. Assuming local thermodynamic equilibrium, it follows that the electron energy distribution is Maxwellian, which yields an electron current given by Eq. \ref{Ie} and Eq. \ref{Ies} \cite{Merlino2007}:
\begin{equation}\label{Ie}
    I_e = I_{esat}\exp\bigg(\frac{e(V_p-V_s)}{k T_e}\bigg)
\end{equation}
\begin{equation}\label{Ies}
    I_{esat} = \frac{1}{4}e n_e A_p \sqrt{\frac{8 k T_e}{\pi m_e}}
\end{equation}
where $n_e$ is the electron density, $e$ is electron charge, $A_p$ is the surface area of the probe, $T_e$ is the electron temperature, and $m_e$ is the electron mass.
A plot of $\ln{I_e}$ vs. $V_p$ produces a straight line below $V_s$, whose slope yields $T_e$. Obtaining $I_e(V_p)$ from the measured $I$-$V_p$ characteristics  requires subtraction of the ion current. This may be done in a variety of ways. The simplest is to assume that the ion saturation current is not dependent on $V_p$. If the electron and ion temperature are assumed to be similar, $I_{isat}$ has the same form the as $I_{esat}$, replacing $m_e$ with the ion mass $M$ and $T_e$ with the ion temperature $T_i$. If $T_i << T_e$, $I_{isat}$ is given by the Bohm current \cite{Cherrington1982}. Using this method, the most negative value in the ion saturation region is typically used as the constant value or a simple line is fit and subtracted from the probe data. Despite its simplicity, this approach has limited value for laser-generated plasmas, because it is only valid for a narrow range of laser fluences, target-probe distances, and laser spot sizes. The so-called orbital-motion-limited (OML) theory for cylindrical probes \cite{Weaver1999, Chen2001} accounts for ion current more accurately across a broader range of laser-plasma conditions, and was therefore chosen in this study. OML theory assumes negligible collisions in the sheath surrounding the probe and as such, the current that the probe collects is limited by the orbital motion of ions around the cylindrical probe. This current is given by \cite{Langmuir1926,Chen2001}:
\begin{equation}\label{IivsVp}
    I_i = A_p n e \frac{\sqrt{2}}{\pi}\bigg(\frac{e(V_s-V_p}{M}\bigg)^{\frac{1}{2}}
\end{equation}
OML theory is most accurate when $\xi \equiv R_p/\lambda_d< 3$, where $R_p$ is the probe radius and $\lambda_d$ is the Debye length. However, it is also known to provide valid estimates for $\xi>3$ and even for high density plasmas. OML theory is also widely used in industry for a variety of plasma conditions \cite{Chen2009}.
To even more accurately correct for the ion current, more complex models such as Allen-Boyd-Reynolds (ABR) or Bernstein-Rabinowitz-Laframboise (BRL) can be used which account for finite plasma sheaths and finite sheaths with orbital motion, respectively.
For $3<\xi<100$, the real plasma density lies between what is predicted from OML and BRL theories, with OML providing an upper limit \cite{Chen2001}. In the present study, a maximum value of $\xi$ w$\approx 140$ was noted across all experimental conditions, so OML is considered to provide a robust estimation of $n_e$ and $T_e$ versus simply subtracting the Bohm current, while avoiding the complexities of ABR or BRL theories.
\begin{figure*}[!th]
    \centering
    \includegraphics[width = \linewidth]{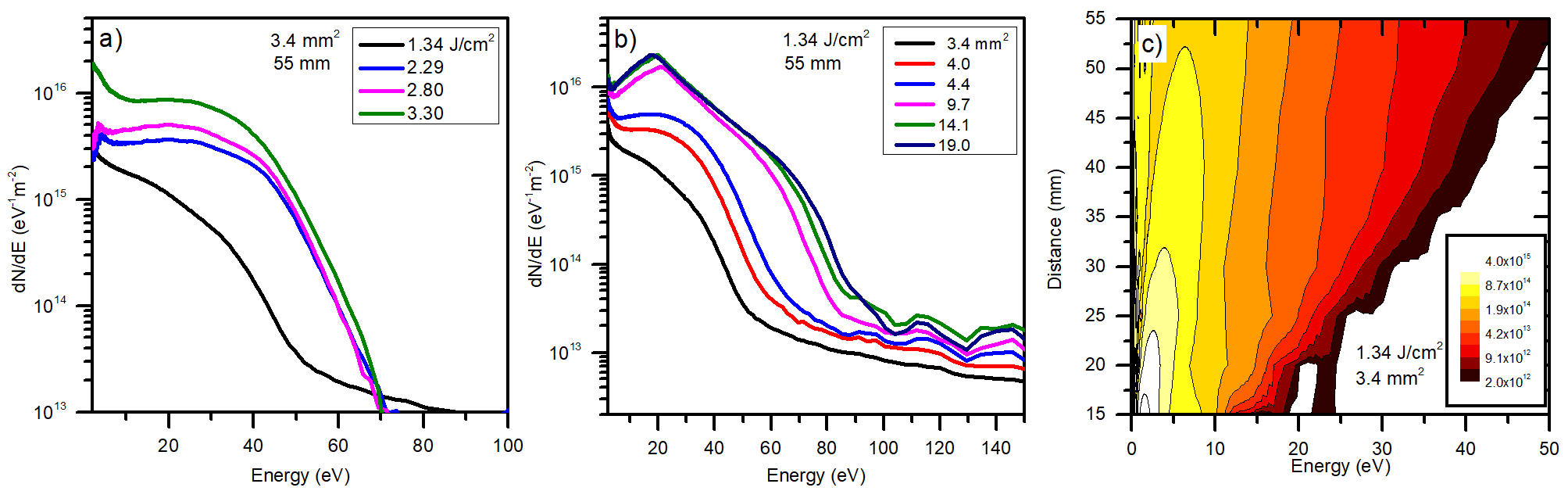}
    \caption{Time of flight (TOF) kinetic energy distribution, taken at 55 mm away from the target for a) various fluences using the same laser spot size of 3.4 mm$^2$, and for b) various spot sizes using the same laser fluence of 1.34 J/cm$^2$. c) Contour map showing the evolution of the TOF kinetic energy distribution as the plasma expands away from the target. Close to the target, the distribution is dominated by a low energy peak $\sim$5 eV, with few higher energy ions out to 22.5 eV. The most probable energy gradually shifts to higher values and the distribution becomes broader as the expansion proceeds, leading to greater relative fractions of high kinetic energy species at long distances from the target. The observed acceleration is attributed to space charge effects with the plasma expansion front exhibiting kinetic energies extending out to 100 eV. Measurements for part c) were performed at 1-mm increments from 15 to 55 mm while ablating the target using a spot size of 3.4 mm$^2$ at 1.34 J/cm$^2$. Values of kinetic energy were evaluated assuming particles with average mass m$_{av}$ between Fe and Se ions.}   
    \label{dNdE}
\end{figure*}

Each $I$-$V_p$ curve is treated in the following way. First, a time is chosen from the $I$-$t$ dataset and the probe current at that time is extracted from all values of probe bias and plotted to construct the $I$-$V_p$ characteristics. Fig. \ref{typical_IV}a shows an $I$-$V_p$ curve on a linear scale before and after ion subtraction, while the inset displays a selection of the $I$-$t$ curves from which it was constructed. Next, $I^2$ vs $V_p$ is plotted and a line is fit whose parameters will be used to determine $n$ (imposing quasineutrality, $n_i\approx n_e\approx n$) and $I_i$ and is shown in Fig. \ref{typical_IV}b.
Finally, $I_i$ is subtracted from the $I$ vs. $V_p$ curve and the slope of the exponential portion of $I_e$ vs. $V_p$ is fit to determine $T_e$ (Fig. \ref{typical_IV}c). The intersection of this line with a line drawn through the electron saturation is used to determine the space potential $V_s$, and this value is used to obtain $n$ employing Eqs. \ref{Ie} and \ref{Ies}. 
Once $n$ and $T_e$ are determined, the Debye length follows from
\begin{equation}\label{debye}
    \lambda_d = \sqrt{\frac{\epsilon_0 k T_e}{n e^2}}
\end{equation}


Less stringent criteria may be applied to interpretation of data from the planar probe. Under strong negative bias, only positive, ionic species
are collected by the planar probe. The ion current measured as
a function of the time delay from the laser pulse can be used
to deduce how the ion density of the plasma changes with
time at the position of the probe \cite{Doggett2009}. The saturation current measured by the planar probe under strong negative bias, $I(t)$, yields the time of flight (TOF) distribution of the ions and, through a change of variables using Eq. \ref{TOFdist}, provide the so-called ``TOF kinetic energy distribution'' of the ions \cite{Franghiadakis1999}, 
\begin{equation}\label{TOFdist}
    \frac{dN}{dE} = \frac{I(t) t^3}{m_{av} e d^2 A_p}
\end{equation}
where $t \equiv$ TOF, $d$ is the distance target-probe distance, $m_{av}$ is the average ion mass, and $A_p$ is the area of the probe.

The TOF kinetic energy distribution 
has long been of particular interest in PLD, as it is well known that low ion kinetic energy accompanied by low ion densities tend to produce crystal growth conditions similar to molecular beam epitaxy with large-grained monolayers possible. Conversely, high kinetic energy of plasma species, leads to
increased depth of implantation and higher density of defects
in grown films.

The TOF data also yields the plasma flow velocity $u$, which may be used in combination with the local speed of sound, $a = (\gamma k T_e/m)^{1/2}$, to evaluate the Mach number $M$ of the plasma fluid expansion,

\begin{equation}\label{Mach_number}
    M \equiv \frac{u}{a} = \frac{u}{(\gamma k T_e/m)^{1/2}}
\end{equation}
where $\gamma = C_p/C_v$, is the ratio of specific heats at constant pressure and constant volume, respectively. Since $M^2$ essentially represents the ratio of the directed kinetic energy of the plasma to the random thermal motion, it is a useful parameter to evaluate how the relative energetic contributions of directed and thermal motion in the plume may affect thin film growth.

\begin{figure*}[t!]
    \centering
    \includegraphics[width=\textwidth]{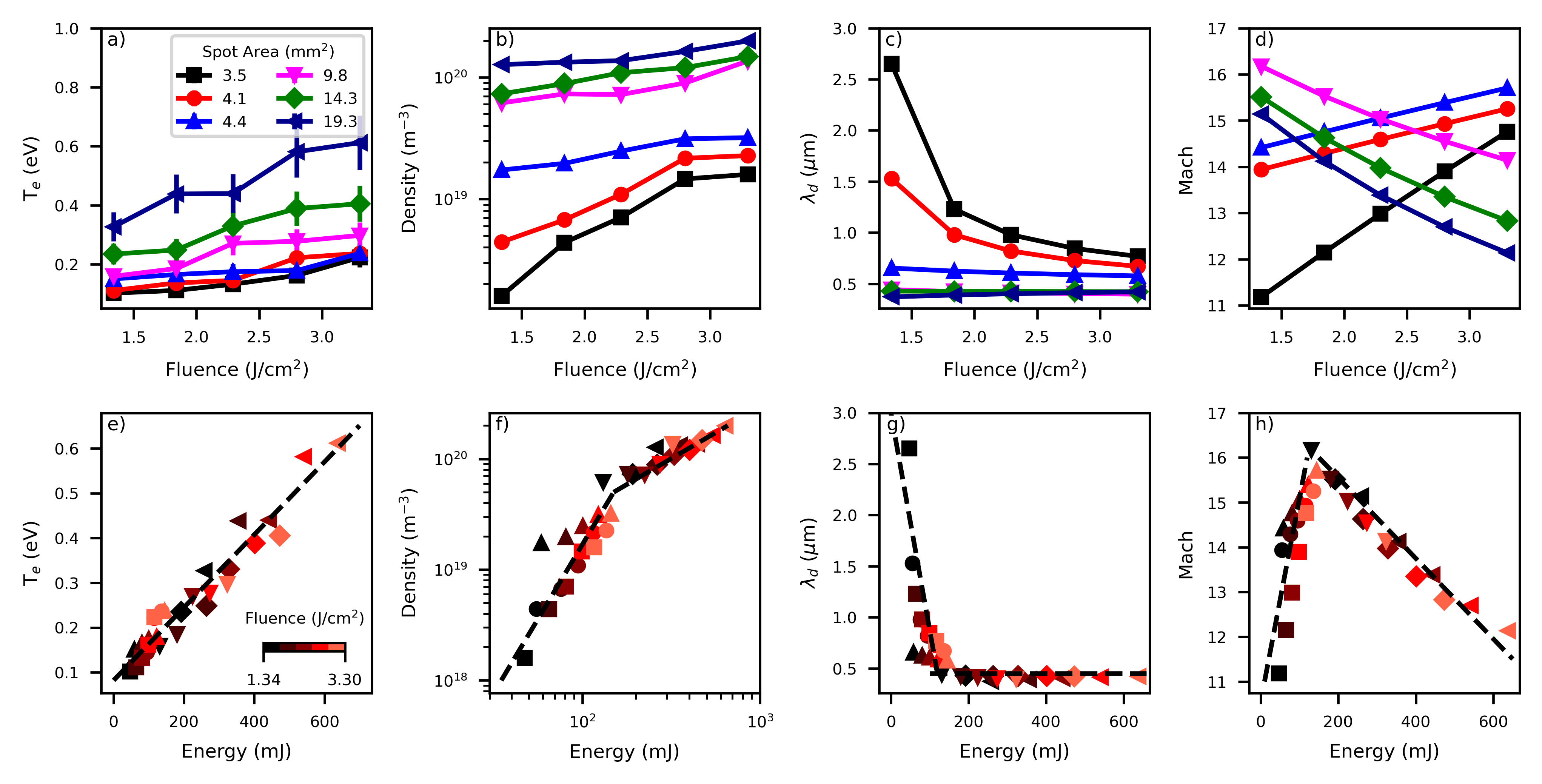}
    \caption{Parameters of the plasma produced from the ablation of the FeSe target, measured at 55 mm from the target surface. The top row shows how the a) electron temperature ($T_e$) b) electron density ($n$), c) Debye length ($\lambda_d$), and Mach number of the expansion frong ($M$) vary with fluence for fixed laser spot sizes. $T_e$ and $n$ both increase monotonically with fluence, for all spot sizes. The bottom row, e-h, displays the same parameters as a function of the pulse energy $E$, which is delivered to the laser-plasma interaction volume and subsequently to the target. While $T_e$ increases linearly with energy, $n$ shows a change in its rate of increase with $E$. The observed shift in the behavior of  $n$ leads to distinct behaviors  in $\lambda_d$. g) $M$ is seen to increase with energy up to $E\approx$180 mJ, and then reverse behavior for greater energies.}  
    \label{probe_results}
\end{figure*}


\section{Results and Discussion}
\label{S:4}
We first consider the TOF kinetic energy distribution of ions produced during FeSe ablation. Fig. \ref{dNdE} shows how the general characteristics of such distribution, measured by the planar Langmuir probe, vary with three of the  most accessible experimental parameters of PLD crystal growth, the laser fluence, the laser spot size, and the target-to-substrate distance. Fig. \ref{dNdE}a shows that for a fixed spot size, the overall number of particles scales with the laser fluence. At a distance of 55 mm from the target and for a spot size of 3.4 mm$^2$, fluences between 1.34 J/cm$^2$ and 3.30 J/cm$^2$ lead to mean kinetic energies in the $\sim$8-30 eV and the high energy tails that extend to between 50 eV and 70 eV. Two factors leading to the high kinetic energy of the directed motion of the particles are laser-target interaction and the absorption of the laser energy by the plasma. Increasing fluence contributes to initial vapors with higher number density and thereby to greater absorption of the laser by such vapor. Despite the complex interplay between these two effects, the overall result of increasing laser fluence is a denser plasma and more energetic expansion. In addition, the increasing width of the kinetic energy distribution is partially indicative of increased plasma heating.

Another long recognized characteristic of PLD is the variation of the angular distribution of plume species with different experimental conditions. This is captured in part by the dependence of the ion kinetic energy distribution on the laser spot size, as seen in the Fig. \ref{dNdE}b. Once again, the kinetic energy distribution broadens with increasing spot size. This behavior, measured here for FeSe ablation, has been observed in numerous other materials and experimental configurations. It has been reported, for example, as a broadening in the full width at half maximum of ion current traces vs. time collected in Faraday cup experiments \cite{Grun1986} and changes in the width of the velocity distribution of ions as a function of laser spot tightness \cite{Harilal2007}. Changes in spot size alter the interaction volume between the laser pulse and the initial vapor, and may also affect the efficiency of evaporation due to heat dissipation effects in the target. Several competing effects here may lead to changes in the geometrical character of the plume, making it transition between highly directed (quasi-1D) expansions to more 3D-like behavior.  
In fact, Fig. \ref{dNdE}b suggests a qualitative difference between plasma expansions produced by small spot sizes ($\lesssim$4.4 mm$^2$) and large ones ($\gtrsim$9.7 mm$^2$). At small spot sizes, the mean of the distribution gradually shifts to higher values while the distribution broadens. For large spot sizes, distribution broadening seems to be the main variation, with a now well-defined distribution peak essentially ``pinned'' at a fixed value. 
In all cases exhibited in Fig. \ref{dNdE}b, the relative fraction of more energetic species increases with spot size, with high energy tails extending to between 50 eV and 100 eV. 
Fig. \ref{dNdE}c shows the evolution of the TOF distribution as the laser plume moves away from the target for a spot size of 3.4 mm$^2$ and a laser fluence  of 1.34 J/cm$^2$. The most probable kinetic energy is observed to shift to higher energy as the plasma expands away from the target. The width of the distribution also increases, leading to greater fractions of high kinetic energy species at distances far from the target. This broadening of the kinetic energy distribution and acceleration of the particles, while the overall plume is actually cooling, is related to space charge acceleration effects, and is another signature characteristic of PLD \cite{Hairapetian1991}. A scenario that qualitatively explains this phenomenon 
is the early ejection of a relatively low-density, highly energetic electron population during the ablation event and/or the plasma during its early formation. The resulting charge imbalance creates an electric field that accelerates the directed motion of the ions during the otherwise adiabatic plasma expansion.

The plasma characteristics inferred from the kinetic energy data in Fig. \ref{dNdE} can be significantly expanded by measurements of the plasma parameters at typical distances from the target where substrate placement is intended for thin film deposition. Fig. \ref{probe_results} shows how the electron temperature ($T_e$), the electron density ($n$), the Debye length ($\lambda_d$), and the Mach number of the expansion front ($M$) vary with laser fluence and laser spot size at 55 mm away from the FeSe ablation target. As described in section \ref{S:3}, the values of $T_e$ were obtained from $I$-$V_p$ curves extracted at the time of maximum ion current, which has previously been shown to be the temporal coordinate that yields maximum electron temperature \cite{Toftmann2000}. 

Fig. \ref{probe_results}a-d show how the plasma parameters depend on laser fluence, for fixed spot areas. The electron temperature increases monotonically with fluence for all spot areas (Fig. \ref{probe_results}a). For a spot area of 3.5 mm$^2$, for example, $T_e$ ranges from 0.1 eV at 1.34 J/cm$^2$ to 0.2 eV at 3.3 J/cm$^2$. These values are similar to the majority of PLD plasmas reported in the literature \cite{Doggett2009}.
Larger spot sizes produce hotter plasmas with $T_e$ as high as $\sim$0.6 eV for a spot area of 19.3 mm$^2$ with a fluence of 3.3 J/cm$^2$. The electron density (Fig. \ref{probe_results}b) shows a similar trend, spanning the  1$\times$10$^{18}$-2$\times$10$^{20}$ m$^{-3}$ range with the highest densities achieved for large spot sizes and high fluences. For spot areas greater than 9.8 mm$^2$, the ratio of $T_e/n$ remains approximately constant as fluence is increased, which is evident by an essentially constant $\lambda_d$ shown in Fig. \ref{probe_results}c with a value of $\sim$0.4 $\mu$m.
For a spot area of $\sim$4.4 mm$^2$, $\lambda_d$ is still approximately constant with laser fluence but with a greater value at $\sim$0.7 $\mu$m. The two smallest spot sizes lead to gradual increase towards greater $\lambda_d$ as fluence decreases, reaching its highest  value of $\sim$2.7 $\mu$m for 1.3 J/cm$^2$. 

The dependence of $M$, the Mach number of the expansion front, with fluence is rather defining. As shown in Fig. \ref{probe_results}d, $M$ increases with fluence for the smaller spot sizes, but exhibits a reversed, monotonically decreasing trend with fluence for the larger spot sizes. This reversal suggests a switch in energy apportionment between the directed and thermal motion components of the expansion.

\begin{figure*}[!th]
    \centering
    \includegraphics[width = \linewidth]{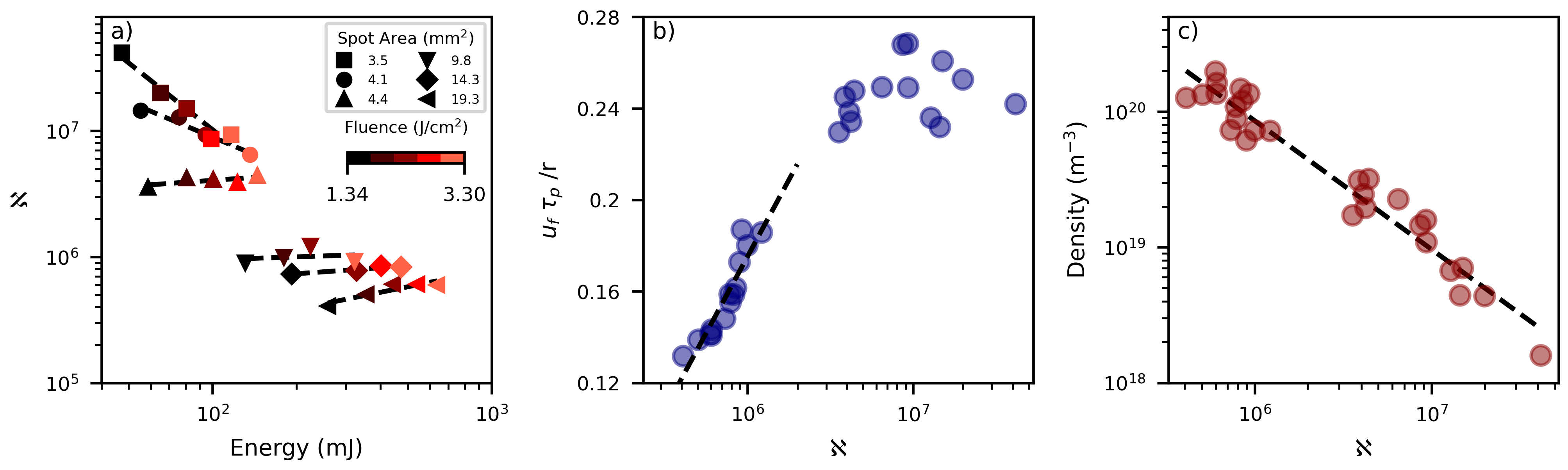}
    \caption{a) Dimensionless parameter $\aleph$ vs. pulse energy, showing that the two plasma regimes are characterized by significantly different number of available photons per absorber in the laser-plasma interaction volume. b) The ratio of the  plume extent during the laser pulse ($u_f \tau_p$) to the spot radius $r$ vs. $\aleph$ indicates that the high energy regime is characterized by a less forward peaking than the low energy plasma. c) Uniform scaling of the electron density with $\aleph$ suggests that a process governed by the number of photons relative to number of absorbers in the laser-plasma interaction volume is what determines  the transition between the two plasma regimes.}
    \label{Aleph}
\end{figure*}

Further clarifying trends become apparent when the plasma parameters are plotted as a function of pulse energy, $E$ = $Fluence$ $\times$ $Spot$ $Area$. While still carrying the spot size information due to simple proportionality, plots in terms of $E$ are more insightful for inferring signatures of laser-plasma and laser-target interactions. As seen in Fig. \ref{probe_results}e-h, the dependence of the plasma parameters on $E$ is rather distinctive. $T_e$ scales linearly with the pulse energy for all fluences and spot areas (Fig. \ref{probe_results}a). The electron density dependence on $E$ shows evidence of two distinct plasma regimes, with a transition in the rate of increase of $n$ with energy at $E\approx$180 mJ. Unlike the threshold for plasma absorption, which occurs at substantially lower energies\cite{Geohegan1994}, plasma absorption is clearly present in both regimes observed in Fig. \ref{probe_results}b, given the high values of $M$, which cannot be supported in absorption-less expansions \cite{Kelly1990,Kelly1992}.

Virtually all plumes associated with $E<180$ mJ correspond to the smaller spot areas for which $M$ scales with fluence. Here, as $n$ is reduced over more than one order of magnitude with reducing spot size, the Debye length is seen to increase significantly (Fig. \ref{probe_results}g), as expected from Eq. \ref{debye}. In the high density regime $E>180$ mJ, $\lambda_d$ is essentially constant with pulse energy and spot size. The impact of the two plasma regimes on how $M$ scales with $E$ is seen in Fig. \ref{probe_results}h. The forward peaking of both plasmas is strong, with high ratios of kinetic to thermal energy ($\propto M^2$) for all values of $E$. However, increase in the directed motion of the expansion is favored over the thermal component for $E<180$ mJ leading to $M$ that increases with $E$, while the opposite is true for $E>180$ mJ.  
 

Measurements were also carried out at a distance of 15 mm from the target. Trends are essentially the same, but the magnitudes of $T_e$ and $n$ are greater. Considering the minimum case of 3.5 mm$^2$ spot area and 1.34 J/cm$^2$ fluence, $T_e = 0.64$ eV and $n$ = 1.66$\times$10$^{19}$ m$^{-3}$ which represent significant increases in density and temperature relative to the same conditions at 55 mm. At 15 mm, only cylindrical probe data collected up to 9.7 mm$^2$ spot area could reliably be analyzed using OML theory, which is evident by a change in the shape of the $I$-$V_p$ characteristic curve. The inset in Fig. \ref{typical_IV}(b) shows two $I$-$V_p$ curves for 2.8 J/cm$^2$ collected at 15 mm from the target. The ion current for a spot area of 4.0 mm$^2$ still follows the $I^2$-$V_p$ relationship predicted by OML, with $T_e$ and $n$ of 1.04 eV and 5.15$\times$10$^{19}$ m$^{-3}$ respectively. With larger spot area, however, the ion current continues to become more negative with increasingly negative bias and can no longer be fitted with OML scaling. As discussed in section \ref{S:3}, a different probe geometry would be needed for studies with Langmuir probes very close to the ablation target. The availability of density and temperature data for two different positions (15 mm and 55 mm from the target) along the central axis of the plasma expansion, allows us to check the consistency of our measurements. Since both positions are well within the adiabatic stage of the plasma expansion, one would expect their states to be related by the well known expression for adiabatic processes,
\begin{equation}\label{isentrope}
    \frac{T_2}{T_1} = \bigg(\frac{n_2}{n_1}\bigg)^{(\gamma-1)}
\end{equation}
where again $\gamma = C_p/C_v$. Using $T_e$ and $n$ for each pair of measurements taken at 15 mm and 55 mm in which OML was applicable (20 pairs), the average ratio of specific heats was determined to be $\gamma = 1.6 \pm 0.11$. This value is close to what is expected of a monatomic ideal gas ($\gamma = 5/3$), which is a widely accepted ansatz in the description of plasma phenomena \cite{Zeldovich2002}, and has been routinely invoked in the study laser plasmas \cite{Anisimov1996}, indicating the internal consistency of our plasma measurements.

The origin of the two plasma regimes identified in Fig. \ref{probe_results} can receive greater elucidation by introducing the dimensionless parameter \begin{equation}\label{Harris_Number}
    \aleph = \frac{E/h \nu}{n A u_f \tau_p}
\end{equation}
where $E$ and $h \nu$ are the pulse and photon energy, respectively, while $A$ stands for the laser spot area, $u_f$ is the velocity of the plasma front, and $\tau_p$ is the pulse duration. 

The parameter $\aleph$ is an estimate of the number of photons delivered to the laser-plasma interaction volume during the pulse ($E/h \nu$) with respect to the number of ionized centers in such interaction volume ($n A u_f \tau_p$). The factor $n A u_f \tau_p$ scales with the total number of absorption centers in the interaction volume, and is a reasonable stand-in for the later in scaling considerations.

\begin{figure*}[t!]
    \centering
    \includegraphics[width=0.9\textwidth]{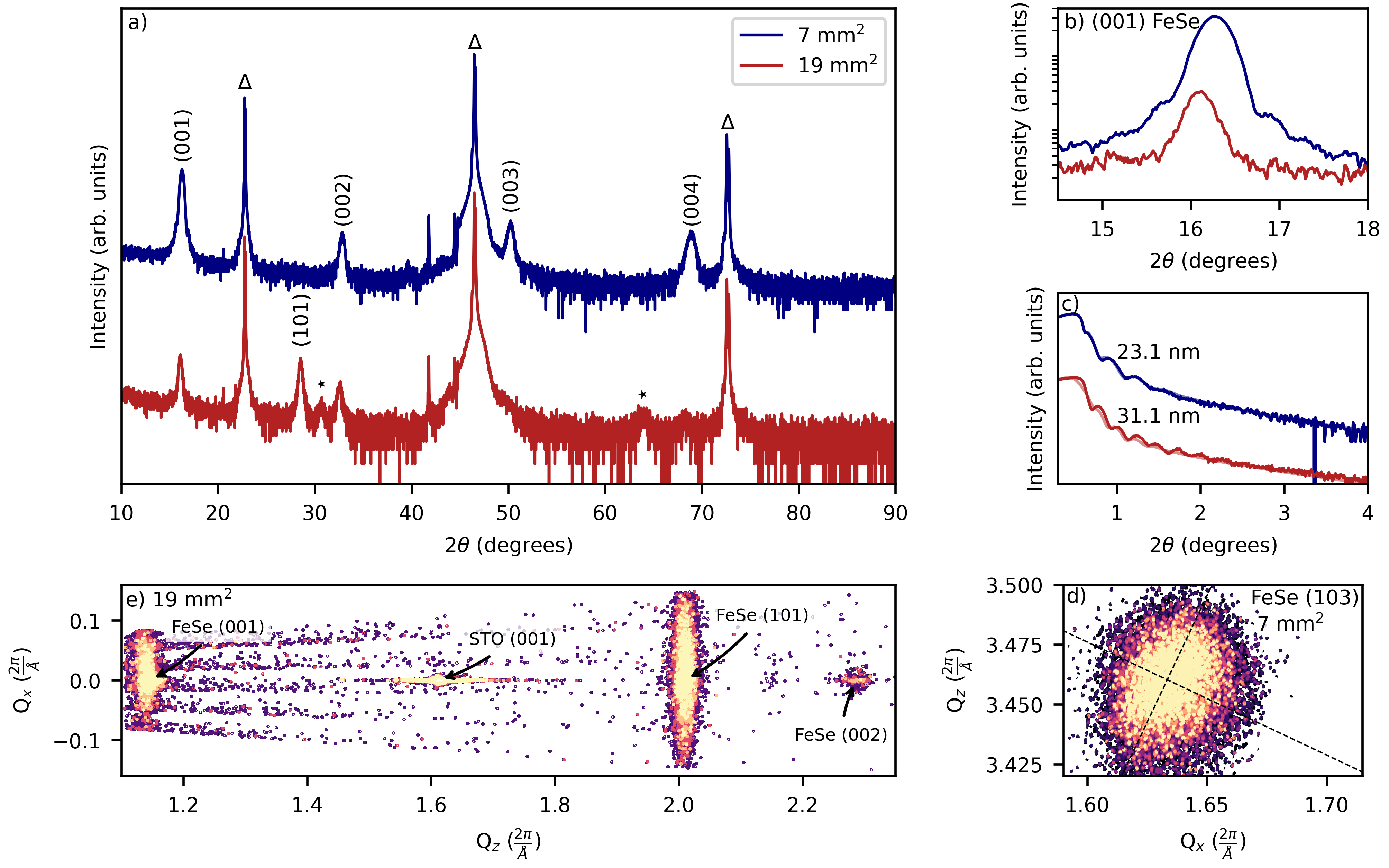}
    \caption{X-ray diffraction of two FeSe films grown on TiO$_2$-terminated SrTiO$_3$ substrates using small and large spot sizes, 7 mm$^2$ and 19 mm$^2$, respectively. a) $\theta$-2$\theta$ scan shows that the film grown with the smaller spot size is $c$-axis oriented FeSe while the film grown with the large spot shows two different orientations, (001)- and (101)-oriented FeSe. b) Detailed view of the FeSe (001) peak shows a clear difference in the $c$ lattice parameter. Thickness fringes are clearly seen on the film with the smaller spot size, indicating high quality. c) X-ray reflectivity was used to determine the thickness of each film by fitting the XRR curve. The growth rate for the 19 mm$^2$ spot is $\sim$10 times greater that the 7 mm$^2$ spot. d) Reciprocal space map of FeSe (103) on the 7mm$^2$ spot size sample. The (103) peak is symmetrical in reciprocal space, indicating minimal compositional and strain gradients, defects, and mosaicity. The (103) peak was too weak to detect in the 19 mm$^2$ spot sample. e) Wide area RSM of the 19 mm$^2$ sample confirms epitaxial orientations of the (001)- and (101)- oriented domains. The (101)-oriented domain is much more broad in the Q$_x$ direction than the (001)-orientation which suggests a greater degree of mosaicity or defects.}  
    \label{XRD}
\end{figure*}

Fig. \ref{Aleph}a shows that the two plasma regimes are characterized by significantly different numbers of available photons per absorbers in the laser-plasma interaction volume. For $E<180$ mJ, substantially more photons are present per absorber, while  this number drops by at least one order of magnitude for $E>180$ mJ. This suggests that a change in laser-plasma interaction $E\approx$180 mJ is responsible for triggering a switch of plasma regimes. According to the temperature data (Fig. \ref{probe_results}e), the plasma above 180 mJ is hotter. This is consistent with the lower values of $u_f \tau_p/r$ noted in Fig. \ref{Aleph}b for $E>180$ mJ. This corresponds to plumes of greater thermal content compared to the lower energy plumes. This is also in agreement with the Mach number trends noted in Fig. \ref{probe_results}d,h. The uniform scaling of the electron density with $\aleph$, with the absence of the rate of increase noted in the energy scaling (Fig. \ref{probe_results}f) suggests that indeed a process governed by the number of available photons relative to the number of absorbers in the laser-plasma interaction volume is what determines the transition between the two plasma regimes.

To study the potential effect of the laser plasma on crystal growth, FeSe thin films were deposited under two different conditions corresponding the two identified plasma regimes. Namely, FeSe was grown using a small and a large spot size, 7 mm$^2$ and 19 mm$^2$ respectively. Figure \ref{XRD} shows the XRD analysis of the two samples. The $\theta$-2$\theta$ scan in Figure \ref{XRD}a shows that each of the films contain epitaxial, c-axis oriented FeSe, evidenced by the presence of the (00$\ell$) family of planes. The sharp peaks marked with a triangle ($\Delta$) are the (001), (002), and (003) planes of the SrTiO$_3$ substrate. While the small aperture sample contains only the c-axis orientation, the large aperture sample is a mixture of two epitaxial orientations, (001)- and (101)-oriented FeSe, as well as some impurity peak that cannot be attributed to FeSe or its other crystal phases (marked with a star $\star$). Figure \ref{XRD}b shows the (001) peak for each sample. The small aperture sample shows clear thickness fringes from dynamical diffraction (Pendell\"{o}sung oscillations) 
which indicates high crystal quality and a sharp interface with the substrate. The shift in peak location also indicates that the c lattice parameters for each sample are different. Calculated from the average lattice constant for each (00$\ell$) reflection, the small aperture sample has c = 5.452 $\pm$0.003 \AA and the large aperture sample has c = 5.500 $\pm$0.005 \AA. The weighted average of $K_{\alpha1}$ and $K_{\alpha2}$ ($K_\alpha$ = 1.54187 \AA) was used for calculations since the separate diffraction peaks are not resolved in the film peaks. Both are shorter than the bulk c parameter of 5.525 \AA. 
The RSM of the (103) plane for the 7mm$^2$ aperture sample is shown in Figure \ref{XRD}d and was used to calculate the in-plane lattice parameter a = 3.839 $\pm$0.003 \AA. The 103 peak is symmetrical, indicates little very little broadening from mosaicity, defects, or composition and strain gradients.
Lastly, a wide area RSM of the large aperture sample shows the broadening of each of the two epitaxial orientations (Figure \ref{XRD}e). The (101)-oriented domain is more broad in Q$_x$, which suggests more mosaicity or defects than the (001)-oriented domain.  The (002) is especially more narrow in comparison to the 101.
The thickness for each film was kept close the the same to eliminate the effects of film thickness on the analysis of crystal quality. Figure \ref{XRD}c shows XRR for each film and the fit that was used to calculate the thickness and roughness of the films. The 7 mm$^2$ aperture film is 23.1 nm thick with a roughness of 1.4 nm and the 19mm$^2$ aperture film is 31.1 nm thick with 2.5 nm roughness.
With these thicknesses, the growth rate for each aperture was determined to be 0.0385 \AA/pulse for the small and 0.415 \AA/pulse for the large aperture. The growth rate per pulse with the large aperture is ~10 times greater than the smaller one. 

These findings can be understood in terms of the classical view of thin film growth during PLD. The pure phase single crystalline epitaxial layers obtained  at 7 mm$^2$ follow from the relatively low kinetic energy of the depositing species, which avoids damage to the (100)-oriented TiO$_2$-terminated STO surface. The high nucleation rate due to high particle flux, results in layer-by-layer growth of high quality epilayers. Films grown at with a spot area of 19 mm$^2$ form under conditions of substantially higher-energy bombardment. Damage to the TiO$_2$-terminated surface likely results in the nucleation of grains with different orientation, that compete in their growth with the (100) orientation. This leads to doubly-epitaxial films exhibiting more than one crystal orientation with well defined epitaxial relationship to to substrate and to each other. 

It is envisioned that for ultrathin layers or depositions at lower temperatures, plasma parameter effects may become noticeable and lead to novel levels of control in the growth of heterostructures.

\section{Conclusion}
\label{S:6}
The plasma generated during PLD of FeSe exhibits rich phenomenology with a variety of well defined trends for the plasma parameters, which define two distinct laser plasma regimes. In one regime, characterized by a high number of photons with respect to absorbers in the laser-plasma interaction volume, increases in laser energy favor the directed motion of the plasma over its heating. In a higher energy regime, the photon/absorber ratio is reduced and increases in energy delivered to the plasma are preferentially apportioned to its thermal component at the expense of the directed motion. Epitaxial thin films grown under these two different plasma regimes exhibit different microstrutures. Although the deviations in epitaxial growth can be explained without direct recourse to the scaling of the plasma parameters, such parameters provide guidance maintaining stable growth conditions and for fine tuning operation. Moreover, epitaxy of ultrathin layers and heterostructures deposited at lower temperatures, may be amenable to direct effects caused by changes in plasma parameters. 


\section*{Acknowledgments}
This work was supported in part by the National Science Foundation (NSF) EPSCoR RII-Track-1 Cooperative Agreement OIA-1655280 and by the National Aeronautics and Space Administration (NASA)-Alabama Space Grant Consortium, Research Experiences for Undergraduates (REU) award to UAB. SBH acknowledges graduate fellowship support from the National Aeronautics and Space Administration (NASA) Alabama Space Grant Consortium (ASGC) under award NNX15AJ18H. Any opinions, findings, and conclusions or recommendations expressed in this material are those of the authors and do not necessarily reflect the views of NSF or NASA.

\bibliographystyle{apsrev}
\bibliography{main.bib}

\end{document}